\documentstyle[11pt,paspconf,epsf]{article}

\begin{document}

\title{Black holes and central surface brightness cusps}

\author{Roeland P.~van der Marel}

\affil{Space Telescope Science Institute, 3700 San Martin Drive,
Baltimore, MD 21218, USA}

\begin{abstract}
HST observations show that the surface brightness profiles of
early-type galaxies have central cusps. I summarize here the results
of van der Marel (1999), which show that the observed characteristics
of these cusps are consistent with the hypothesis that all early-type
galaxies have central black holes (BHs) that grew adiabatically in
homogeneous isothermal cores. The models suggest a roughly linear
correlation between BH mass and $V$-band galaxy luminosity, $\log
M_{\bullet} \approx -1.83 + \log L$ in solar units, similar to the
relation suggested by kinematical BH detections in nearby galaxies.
\end{abstract}

\keywords{black holes, surface photometry, cusps, galactic nuclei}

\section{Introduction}

The high spatial resolution of the Hubble Space Telescope (HST) has
allowed astronomers to study the photometric structure of galactic
nuclei with unprecedented detail. Most studies have focused on
early-type galaxies and bulges, and many systems have now been imaged
(e.g., Lauer et al.~1995; Carollo et al.~1997). The main result is
that at the $\sim 0.1''$ resolution limit of HST, virtually all
galaxies have surface brightness cusps, $I \propto r^{-\gamma}$, with
$\gamma>0$ and no observed transition to a homogeneous core. The
observed profiles fall in two categories: `core' profiles, which have
a break at a resolved radius and a shallow slope inside that radius
($\gamma \leq 0.3$), and `power-law' profiles, which have no clear
break but maintain a steep slope ($\gamma > 0.5$) down to the
resolution limit (Faber et al.~1997). Galaxies with $M_V < -22$ have
core profiles, galaxies with $M_V > -20.5$ have power-law profiles,
and both profile types occur in galaxies with $-22 < M_V <
-20.5$. These observations pose a variety of questions, in particular:
what is the origin of the cusps? why is there a range of cusp slopes?
and why does the cusp slope correlate with luminosity?

Here I address to what extent the observations can be explained by a
scenario in which all galaxies have BHs that grew adiabatically into
pre-existing homogeneous cores. This scenario was first proposed and
studied by Young (1980). Although it is not the only available
scenario to explain the presence of surface brightness cusps in
galaxies, its successful application to M87 and M32 (Young et
al.~1978; Lauer et al.~1992a,b) certainly makes it one of the more
attractive ones. Figure~1 shows the surface brightness profiles that
result from adiabatic growth of BHs of various masses $M_{\bullet}$
into an isothermal sphere model. The BH growth yields the well-known
asymptotic slope $I \propto r^{-1/2}$ at small radii. However, at the
radii observable with HST approximately $I \propto r^{-\gamma}$; the
observed slope $\gamma$ increases monotonically with $\mu \equiv
M_{\bullet} / M_{\rm core}$, where $M_{\rm core} \equiv {4\over 3}\pi
\rho_0 r_0^3$ is a measure of the mass of the initial isothermal core.

\begin{figure}
\epsfxsize=0.6\hsize
\centerline{\epsfbox{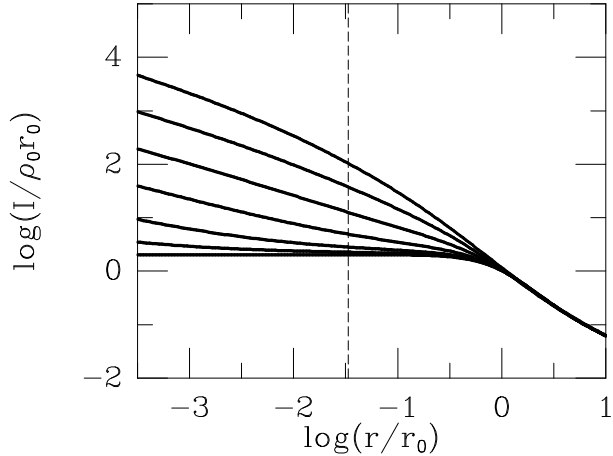}}
\caption{Predicted intensity profiles $I(r)$ for models of adiabatic
BH growth in an isothermal sphere, for models with $\mu$ ranging from
$0$ to $3.3$. The dashed vertical line marks the resolution limit for
a typical HST observation. The quantities $r_0$ and $\rho_0$ are the
core radius and central density of the initial model.}
\end{figure}

\section{Predictions based on scaling relations}

To obtain predictions for the surface brightness profiles of galaxies
from Young's models one must know both the black hole mass and the
characteristics of the homogeneous progenitor cores. These quantities
all scale with galaxy luminosity. Kinematical BH detections in nearby
galaxies indicate an approximately linear relationship between
$M_{\bullet}$ and galaxy (spheroid) luminosity $L$ (e.g., Kormendy \&
Richstone 1995; Magorrian et al.~1998; see also Figure~3 below). For
core galaxies, the observed nuclear photometric parameters obey
scaling relations similar to those of the fundamental plane (Faber et
al.~1997). In the scenario studied here, these photometric relations
must reflect properties of the initially homogeneous cores from which
the present cusps formed. It is tempting to assume that the
homogeneous progenitor cores of power-law galaxies followed the same
relations, because core galaxies and power-law galaxies also follow
the same global fundamental plane relations. 

With these arguments and relations, Young's models predict a unique
surface brightness profile as function of luminosity. Figure~2 shows
the results at the distance of Virgo. The dimensionless BH mass scales
as $\mu \propto L^{-0.5}$, because $M_{\bullet} \propto L$ and $M_{\rm
core} \propto L^{1.5}$. Hence, steeper cusps are predicted in
lower-luminosity galaxies. Figure~2 shows that core profiles are
predicted for $M_V < -21.2$, and power-law profiles for $M_V > -21.2$.
This reproduces both the sense and the absolute magnitude of the
observed transition. Intrinsic scatter in BH and galaxy properties can
explain why both types of galaxies are observed around the transition
magnitude.
 
\section{The black hole mass distribution} 

These results suggest that the observed surface brightness cusps in
galactic nuclei are consistent with the typical BH masses determined
kinematically. This can be verified in more detail by fitting Young's
models to the HST surface brightness profiles of individual galaxies.
I have done this for a sample of 46 galaxies with published HST
photometry and $M_V <-19.7$ (for less-luminous galaxies one expects to
resolve neither the clear signature of a core, nor that of a BH, at
the $0.1''$ resolution of HST). The fits are generally quite good,
with typical RMS residuals of 0.03--0.08 mag/arcsec$^2$.

Figure~3a shows the BH masses thus inferred as function of galaxy
luminosity. There is a clear correlation. The best-fit line of unit
slope is $\log M_{\bullet} \approx -1.83 + \log L_V$ (solar
units). This is fully consistent with the correlation inferred from
kinematical BH detections in nearby galaxies (Figure~3b), $\log
M_{\bullet} \approx -1.96 + \log L_V$. Also, photometrically and
kinematically determined BH masses agree to within $\sim 0.25$ dex RMS
for galaxies that have both.

\begin{figure}
\epsfxsize=0.8\hsize
\centerline{\epsfbox{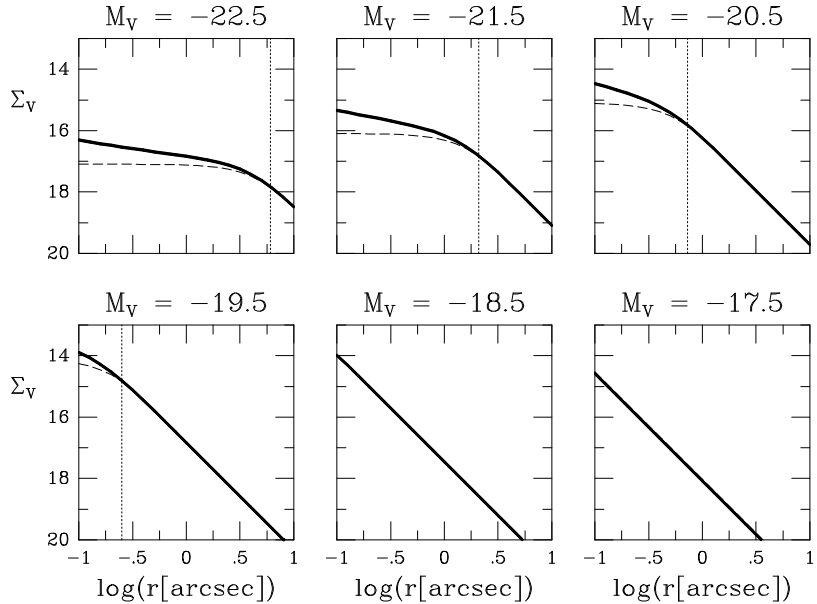}}
\caption{Predicted $V$-band surface brightness profiles
$\Sigma_V(r)$ for galaxies of different absolute magnitude at the
distance of the Virgo cluster, based on Young's models of adiabatic BH
growth and established scaling relations. Dashed curves show the model
profiles before BH growth; heavy solid curves show the model profiles
after BH growth. A dotted vertical line in each panel indicates the
core radius $r_0$ of the initial model.}
\end{figure}

These results provide additional support to the hypothesis that every
galaxy has a central BH, and that the BH mass correlates with galaxy
(spheroid) luminosity. The correlations in Figure~3 are consistent
with quasar statistics, if every galaxy spheroid harbors a BH that was
formed in a quasar phase through matter accretion with efficiency
$\epsilon \approx 0.04$ (van der Marel 1998).

All this demonstrates that Young's photometric models provide a useful
estimate of BH masses in cases where high spatial resolution
kinematical data are unavailable. This is somewhat remarkable, since
the models ignore, among other things, the influence of
mergers. Further work is required to determine what these results
teach us about the physics and timescale of BH formation.

\acknowledgments

I thank Gerry Quinlan for kindly allowing me to use his adiabatic BH
growth software. This work was supported by an STScI Fellowship. STScI
is operated by AURA Inc., under NASA contract NAS5-26555.

\begin{figure}
\centerline{\epsfysize=0.4\hsize
              \epsfbox{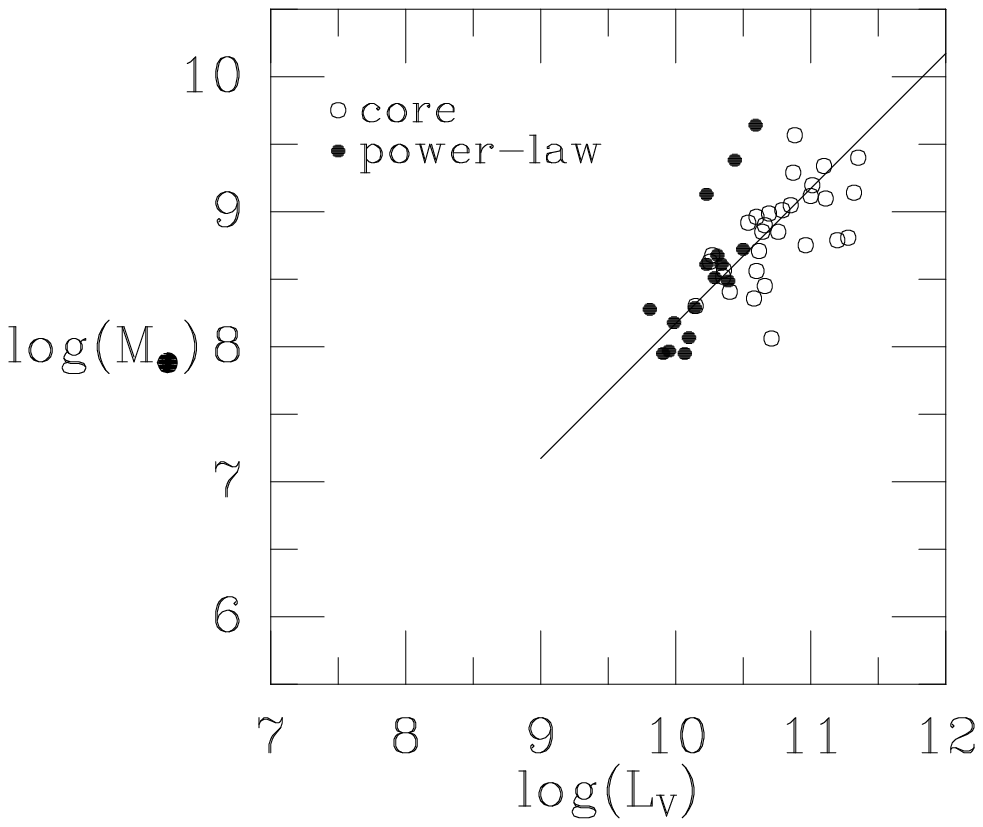}\qquad
            \epsfysize=0.4\hsize
              \epsfbox{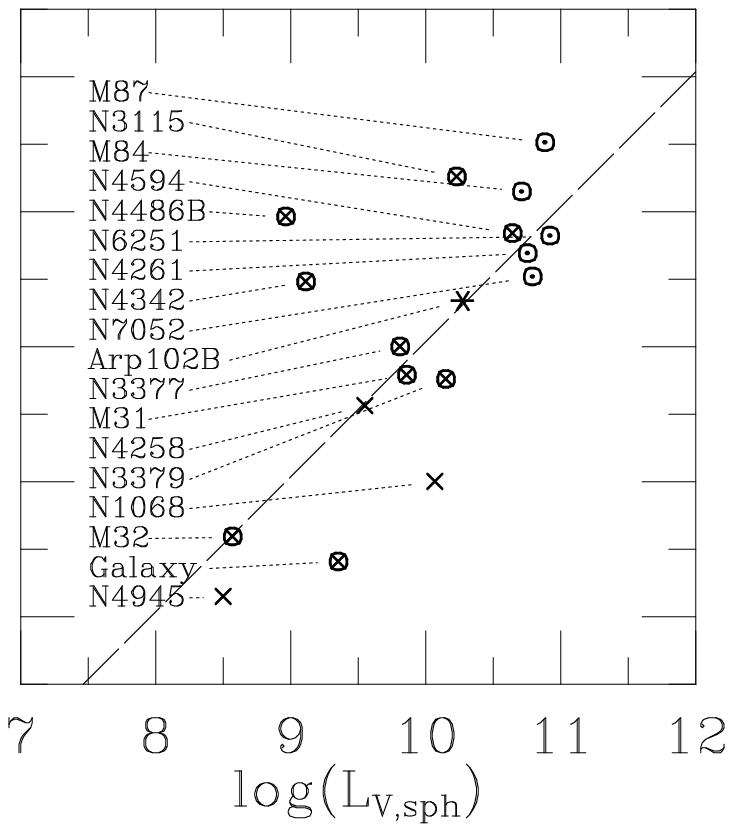}}
\caption{(a; left) BH mass distribution as function of $V$-band
galaxy luminosity, inferred from adiabatic BH growth models for
galaxies with published HST photometry. The solid line is the best fit
line of unit slope. (b; right) BH mass distribution as function of
$V$-band spheroid luminosity for nearby galaxies with
kinematically detected BHs (adapted from van der Marel \& van den
Bosch 1998). The long-dashed line is the best fit line of unit slope,
and is is consistent with the solid line in the left panel to within
$0.13$ dex.}
\end{figure}


\begin{references}

\reference Carollo, C.~M., Franx, M., Illingworth, G.~D., Forbes D.~A. 
           1997, ApJ, 481, 710

\reference Faber, S. M., et al. 1997, AJ, 114, 1771

\reference Kormendy, J., \& Richstone, D. 1995, ARA\&A, 33, 581 

\reference Lauer, T.~R., et al. 1992a, AJ, 103, 703
 
\reference Lauer, T.~R., et al. 1992b, AJ, 104, 552
 
\reference Magorrian, J., et al. 1998, AJ, 115, 2285

\reference van der Marel, R.~P. 1998, in Proc. IAU Symposium 186
           [astro-ph/9712076]

\reference van der Marel, R.~P., \& van den Bosch, F.~C. 1998, AJ, 116, in 
press

\reference van der Marel, R.~P. 1999, AJ, in press [astro-ph/9806365]

\reference Young, P. 1980, ApJ, 242, 1232
 
\reference Young, P., et al. 1978, ApJ, 221, 721

\end{references}
\end{document}